\documentclass{llncs}

\usepackage{amsmath}
\usepackage{amsfonts}
\usepackage{amssymb}
\usepackage{graphpap}
\usepackage{graphics}

\usepackage[normalem]{ulem}
\usepackage{array}
\usepackage{enumerate}
\usepackage{framed}

\usepackage{enumitem}
\usepackage{float}
\usepackage{mathrsfs}
\usepackage{etoolbox}
\usepackage{graphicx}
\usepackage[pdfborder={0 0 0}]{hyperref}
\usepackage{wrapfig}
\usepackage{xcolor}
\usepackage{multirow}

\usepackage{pbox}
\usepackage{hyperref}
\usepackage{mathtools}
\usepackage[hang,flushmargin]{footmisc}
\usepackage{booktabs}

\usepackage{changepage} 

\newcommand{\hide}[1]{}

\newcommand{\esm}[1]{\ensuremath{#1}}

\newcommand{\ms}[1]{\esm{\mathsf{#1}}}
\newcommand{\mi}[1]{\esm{\mathit{#1}}}

\newcommand{\mathsc}[1]{{\normalfont \textsc{#1}}}
\newcommand{\msc}[1]{\esm{\mathsc{#1}}}

\DeclareMathAlphabet{\mathcalalt}{OMS}{zplm}{m}{n}
\newcommand{\mc}[1]{\esm{\mathcalalt{#1}}}

\newcommand{\stitle}[1]{\vspace{5pt} \noindent\textbf{#1.}\ }

\newcommand\mw[1]{\ensuremath{\textcolor{black}{\mathit{#1}}}}
\newcommand{\mydefhead}[3]{\mw{#1} & \mw{#2} & \hfill \mbox{#3} \\}
\newcommand{\mydefcase}[2]{\mid & #1 & \hfill \mbox{#2} \\}

\newcommand{\synteq}{::=}
\newcommand{\emptylist}{\hbox{\tt empty}}

\mathchardef\mhyphen="2D

\newcommand{\sep}{\;\hbox{\tt \slash}\;}
\newcommand{\lsep}{\hbox{\tt:}}
\newcommand{\la}{\langle}
\newcommand{\ra}{\rangle}

\newcommand{\certcon}[1]{\msc{c}\langle #1 \rangle}
\newcommand{\tpccon}[1]{\msc{t}\langle #1 \rangle}
\newcommand{\discon}[1]{\msc{d}\langle #1 \rangle}

\newcommand{\loc}{\mi{loc}}
\newcommand{\bn}{\mi{bn}}
\newcommand{\bp}{\mi{p}}
\newcommand{\lbp}{\mi{P}}
\newcommand{\cert}{\mi{C}}

\newcommand{\pk}{\mi{pk}}
\newcommand{\sk}{\mi{sk}}
\newcommand{\bl}{\mi{B}}
\newcommand{\clist}{\mi{clist}}
\newcommand{\cav}{\mi{c}}

\newcommand{\fpcav}{\mi{fc}}
\newcommand{\tpcav}{\mi{tc}}
\newcommand{\dis}{\mi{d}}
\newcommand{\nonce}{\mi{nonce}}
\newcommand{\sig}{\mi{sig}}
\newcommand\msg{\mi{msg}}

\newcommand\roots{\mc{R}}
\newcommand\con{\mc{C}}

\newcommand\timecav{\ms{timeCaveat}}
\newcommand\targetcav{\ms{targetCaveat}}

\newcommand\fname{\ms{nm}}
\newcommand\fpubkey{\ms{pk}}
\newcommand\fcavs{\ms{cavs}}

\newcommand\fdis{\ms{dis}}
\newcommand\froot{\ms{root}}
\newcommand\hash{\ms{hash}}

\newcommand\sign{\ms{sign}}
\newcommand\verifysig{\ms{verify}}
\newcommand\verifycerts{\ms{IsValidChain}}
\newcommand\validatebl{\ms{IsValidBlessing}}
\newcommand\isrec{\ms{IsRecognized}}

\newcommand\validatefpcav{\ms{IsValidFCav}}
\newcommand\validatecavs{\ms{IsValidCavs}}
\newcommand\bless{\ms{Bless}}
\newcommand\mintdischarge{\ms{MintDischarge}}
\newcommand\fmeaning{\ms{Meaning}_\rho}
\newcommand\isauthorized{\ms{IsAuthorized}}

\newcommand{\prerror}{\hbox{\tt error}}
\newcommand{\prfalse}{\hbox{\tt false}}
\newcommand{\prtrue}{\hbox{\tt true}}
\newcommand{\prcase}{\hbox{\tt case\ }}

\newcommand{\prof}{\hbox{\tt \ of}}
\newcommand{\prthen}{:}

\newcommand{\bprefix}{\preceq}

\newcommand{\allows}{\hbox{\tt  Allow}}
\newcommand{\denys}{\hbox{\tt Deny}}
\newcommand{\allow}[1]{{\allows}\ #1}
\newcommand{\deny}[1]{{\denys}\ #1}
\newcommand{\grp}[1]{#1_G}

\newcommand{\acl}{\mi{A}}
\newcommand{\bdollar}{eob}
\newcommand{\eqdef}{=_{{\rm def}}}
\newcommand{\Undersymbol}{\Downarrow}
\newcommand{\Oversymbol}{\Uparrow}
\newcommand{\rhounder}{\rho_\Undersymbol}
\newcommand{\rhoover}{\rho_\Oversymbol}

\begin{document}
\title{Distributed Authorization in Vanadium}
\author{Andres Erbsen~\inst{1}, Asim Shankar~\inst{2}, and Ankur Taly~\inst{2}}
\institute{MIT, Cambridge, Massachusetts, USA \and Google Inc., Mountain View, California, USA} 
\maketitle

\begin{abstract}
In this tutorial, we present an authorization model for distributed systems that operate
with limited internet connectivity. Reliable internet access remains a luxury
for a majority of the world's population. Even for those who can afford it, a
dependence on internet connectivity may lead to sub-optimal user experiences.
With a focus on decentralized deployment, we present an authorization
model that is suitable for scenarios where devices right next to each other
(such as a sensor or a friend's phone) should be able to communicate securely
in a peer-to-peer manner. The model has been deployed as part of an open-source
distributed application framework called \emph{Vanadium}.
As part of this tutorial, we survey some of the key
ideas and techniques used in distributed authorization, and explain how they
are combined in the design of our model.
\end{abstract}

\section{Introduction}\label{sec:intro}
Authorization is a fundamental problem in computer security
that deals with whether a request to access a resource
must be granted. The decision is  made by a reference monitor
guarding the resource.
Authorization is fairly straightforward in closed systems where
all resources of interest are held on a small set of devices,
and reference monitors have pre-existing relationships
with all authorized principals. In these systems, authorizing a request
involves identifying the principal making the request, and then
verifying that this identity is allowed by the resource's access control
policy. The former is called \emph{authentication}, and the latter
is called \emph{access control}.

Authorization in distributed systems is significantly more complex as
the resources are spread across a network of devices under different
administrative domains~\cite{LABW91}.
Moreover, not all devices and principals in the system may  know each other
beforehand, making even authentication complicated.  For instance, consider the
fairly common scenario of a user Alice trying to play a movie from her internet
video service on her television (TV).  It involves the TV authorizing the
request from Alice to play a movie, and the video service authorizing the
request from the TV to access Alice's account. The video service may
recognize only Alice, and not her TV. The TV must convince the video service that
it is acting on Alice's behalf.

With the advent of the Internet of Things (IoT), various physical devices that
we commonly interact with in our daily lives are controllable over the network,
 and are thus part of a large distributed system. These devices range from tiny
 embedded devices, to wearables, to large home appliances, and automobiles.
 The promise of IoT lies in
multiple devices interacting with each other to accomplish complex tasks for
the user.  For instance, a home security system may interact with security
cameras and locks around the house to ensure that the house is protected from
intruders at all times, and all suspicious activity is logged on the user's
storage server.  Securely accomplishing such tasks involves making several
authorization decisions.  Some of the key questions that arise are \emph{how do
devices identify each other during any interaction?}, \emph{how do users
authorize devices to act on their behalf?}, \emph{how are access control
policies defined?}

Distributed authorization is a long-standing area of research, and several
mechanisms have been designed for a variety of settings. However, most IoT
devices still rely on rudimentary and fragile mechanisms based on passwords and
unguessable IP addresses~\cite{baby-monitor-security,hp-iot-security}. Indeed,
over the last few months, there have been several vulnerabilities and attacks reported on
various ``smart'' devices~\cite{fridge-attack,smart-meter-attack,jeep-attack,wired-iot-report}.
 A study~\cite{hp-iot-security} conducted  by HP on the security of several existing IoT
devices reported  ``insufficient authorization'' as one of the top security concerns.
Besides this, the study found that many devices rely on a service in the cloud for authorization. We stress that
proper authorization is paramount in the IoT setting as authorization breaches
can impact physical security and safety. At the same time, dependence on internet
connectivity can sometimes render ``smart'' devices unusable. Imagine the consequences of
unauthorized access to an embedded heart rate monitor, or being unable to unlock
the door right in front of you due to lack of internet access.

This tutorial explores the design of an authorization model for large, open distributed systems
such as IoT.
The primary guiding
principles behind the design of the model are \emph{decentralized deployment} and
\emph{peer-to-peer communication}. The model does not rely on
any special global authorities for brokering interactions between devices that have a network
path to each other. The justification for these principles is manyfold.
First, a centralized model assumes all entities implicitly trust the default global authorities. This
assumption fails for the enterprise and IoT settings where it is desirable to carve out isolated and fully
autonomous administrative domains for devices. For instance, a user may want to be the sole authority
on all devices in her home. Similarly, an enterprise may want to maintain full control of its devices with
no dependence on any external authority. Second, the central nodes in the system become an attractive
target for compromise to attackers. Given the frequent security breaches at well-reputed
organizations,
users are justifiably weary of having third-party services store more personal data than strictly necessary.
Besides this, protecting personal user data from external and internal threats is quite burdensome for the
organizations as well.
Finally, a centralized model requires that all devices maintain connectivity to
external global services, which is infeasible for many IoT devices, including
ones that communicate only over Bluetooth, or ones present in public spaces
such as shopping malls, buses, and trains, where internet access is unreliable.
Moreover, reliable internet access remains a luxury for a majority of the
world's population, where routing most interactions through a cloud service can
be expensive or simply not possible.

The authorization model presented in this tutorial is fully decentralized, and is based on a distributed
public-key infrastructure similar to SDSI~\cite{RL96}. The model supports peer-to-peer delegation
of authority under fine-grained caveats~\cite{BPETVL14}. Access control policies are based on
human-readable names, and have support for negative clauses and group-based checks. This general
purpose model is applicable in various distributed system settings, including, peer-to-peer computing
environments, IoT, cloud, and  enterprise. The model has been deployed as part of an open-source
application framework called \emph{Vanadium}~\cite{Vanadium}, and is thus referred to as the
Vanadium authorization model. 
As part of this tutorial, we survey a number of key ideas and techniques from previous work on 
distributed authorization, including SPKI/SDSI~\cite{RL96,EFLRTY99}, Macaroons~\cite{BPETVL14},
and the vast body of work on trust management~\cite{BFL96}. We explain how these techniques
are combined by the Vanadium authorization model.

\stitle{Organization} The rest of this tutorial is organized as follows. Section~\ref{sec:desired} describes
the features we desire, and Section~\ref{sec:ideas} surveys key
technical ideas involved in the design of our model. Section~\ref{sec:vanadium} describes our
model in detail, followed by an application of our model to a physical lock device in 
Section~\ref{sec:lock}. Section~\ref{sec:conclusion} concludes.

\section{Desired Features}\label{sec:desired}
In this section, we describe features we seek from our authorization model.
Many of these features have been considered by prior work on
distributed authorization.
We use the term ``principal" informally to refer to an entity in our system,
including, users, devices, processes, and objects, and leave the formal definition
to Section~\ref{sec:vanadium}.

\stitle{Decentralization}
As discussed in the introduction, decentralized deployment and peer-to-peer communication
by default are the central guiding principles behind this work. The model must not force dependence
on special global authorities such as x.509 certification authorities, default identity providers, and proxies
that mediate interactions between principals. Instead, we seek egalitarian systems where \emph{any}
principal can be an authority for some set of other principals. For instance, a user Alice may become an
authority on the identities and access controls on all her home devices. The devices may be configured to
specifically trust only Alice's credentials. 
In general, devices must be able to securely communicate with each other as long as there is a direct
communication channel between them. We seek a model that minimizes interaction with globally
accessible services,
and maximizes what can be achieved with direct peer-to-peer communication.

\stitle{Mutual authentication and authorization}
We require all interactions between between principals to be \emph{mutually}
authenticated and authorized. The principal at each end of a communication channel
must identify the principal at the other end (authentication), and verify
that it is valid in the context of the communication (authorization).
Mutual authentication is essential for both ends to audit all access; we discuss the
benefit of auditing later in this section. Mutual authorization is essential whenever the
communicating principals are mutually suspicious. Unidirectional authorization often
opens the door to rogue entities, leading to 
security and privacy attacks.  For example, when Bob tries to unlock the lock
on Alice's front door, the lock must be convinced that it is communicating with
Bob, and that Bob is authorized to unlock the door. At the same time, Bob must
be convinced that it is indeed sending the request to Alice's front door, and
not an imposter device that is tracking Bob's behavior. 


\stitle{Compound identities}
We live in a world today where all users and devices carry multiple identities. For instance,
a user may have an identity from social media sites (e.g., Facebook identity), an identity from
her work place, an identity from the government (e.g., passport, driver's license), and so on.
Similarly, a device would have an identity from the manufacturer (e.g., Samsung TV model 123),
and an identity from the device owner (e.g., Alice's TV). A principal must be able to act under one
or more identities associated with it, and different identities may grant different authorizations
to the principal.  The authorization model must seamlessly capture this compound nature of a
principal's identity.

\stitle{Fine-grained delegation}
The strength of distributed systems lies in multiple computing agents coming together to accomplish
complex tasks. This is indeed the promise of IoT. For example, Bob would like to play a movie from his
internet video service on Alice's TV and speaker system. In order to enable such interactions, the
authorization model must support flexible sharing and delegation.\footnote{Some systems choose to
distinguish the concepts of ``sharing'' an ``delegation'' with the former being a mechanism for a principal
to allow another principal to access an object while the latter being a mechanism for allowing another
principal to act on its behalf. In this tutorial, we do not make this distinction, and treat ``delegation'' more
broadly as a mechanism for one principal to delegate some of its authority to another principal.}
Alice must be able to delegate access to her TV to Bob, and Bob must be able to delegate access to his
internet video service to Alice's TV for playing a particular movie. 
The model must also support delegations across multiple hops. For instance, Bob must be able to easily
delegate access to Alice's TV to his friend Carol. Moreover, in light of
the decentralization requirement, we would like delegations to work peer-to-peer rather than be mediated
by a central authority.

In practice, delegation of authority is seldom unconditional, and thus the delegation mechanism must support
constraints on the scope of the delegation. For instance, Bob may want Alice's TV to have 
access to his internet video service only for playing a particular movie, and only while Bob is present in Alice's house.
The TV must loose access as soon as Bob leaves the house. Alice may want the same for Bob's access to her TV. 

We emphasize that the delegation mechanism must  be flexible and convenient to use. Inflexibilities or inconveniences
in the mechanism not only affect the user experience but are also detrimental to security as they push users
to look for insecure workarounds. For instance, in the absence of a convenient mechanism to share access to an
internet video service, Bob may end up sharing his account password with the TV, and as a result give
away access to all his account data (e.g., viewing history, purchases) instead of just access to a particular movie.

\stitle{Auditable access}
In a system with delegation, users should be able to audit the use of delegated access
over time. For instance, a user must be able to determine who has access to her devices
and who has \emph{exercised} that access. Delegations are ultimately tied to the intention
of the human end-user, and software must acknowledge that it is impossible to codify all possible
human intentions. This is particularly true when users themselves may be unable to clearly
articulate their intentions at the time of delegation.  An accurate audit trail is a
requirement to detect mismatches between user intentions and their
codification.  For example, Alice might give Charlie access to her home to come
by and walk her dog once a day. Alice's intent is for Charlie to be a dog walker
but she cannot possibly know apriori what time Charlie will come by in all future
days. Auditing Charlie's use of the authority granted to her by Alice is
necessary to detect a violation of the contract between the two. Moreover,
this auditing must be fine grained---if Charlie was tricked into
running a malicious application on her phone, Alice must be able to pinpoint
the exact application that was improperly using the authority she granted to
Charlie.

\stitle{Revocation}
Users make mistakes, devices get stolen/compromised, and relationships break. As a result an
authorization model must support revocation. For instance, Alice must be able revoke Dave's access
to all her home devices when they have a falling out. Similarly, she should be able to revoke all
delegations that she made to her tablet when her tablet gets stolen.

\stitle{Ease of use}
Usability is a key determining factor in the effectiveness of
security systems~\cite{WT99,BJ06}. Systems with complex interfaces and mechanisms
often have degraded security, as users tend to look for insecure workarounds when dealing with them. Thus
we strive to design an authorization model that can be easily understood and configured by system developers,
and lends itself well to simple and clear user interfaces.


\section{Background}\label{sec:ideas}
Distributed authorization is a very mature field with decades of
prior research. The Vanadium authorization model is a result of
combining various known techniques in order to meet the requirements
stated in the previous section. In this section, we provide some background
on distributed authorization, and discuss the key ideas involved in the design
of the Vanadium authorization model.

%

In essence, most authorization mechanisms involve a requester presenting
a set of credentials (possibly obtained from multiple parties) to a reference
monitor guarding a resource which then authorizes the request after validating
the credentials~\cite{S13}. 
A common paradigm is for the requester to present credentials that
establish its identity, which is then checked against an access control policy
(e.g., an access control list (ACL)) by the reference monitor.
Various mechanisms differ in the type of credentials involved. In mechanisms such as
OAuth2~\cite{H12}, OpenID~\cite{openid}, Macaroons~\cite{BPETVL14}, and many
others~\cite{N93,G89}, the credentials are essentially tokens constructed using
symmetric-key cryptography by an issuer (e.g., an identity provider in the case of OAuth2). 
These mechanisms are simple, efficient, easy to deploy, and are widely in
use (particularly OAuth2) for client authorization on the Web. The downside is that the
credentials can be validated and interpreted only by the credential issuer. As a result, the
credential issuer must be invoked for validating credentials during each request\footnote{An
alternative is for each resource owner to become a credential issuer but that leads to a
proliferation of credentials at the requester's end.}. This is undesirable in our setting.

In contrast, mechanisms based on public-key certificates 
\cite{Z95,BFL96,BFI99,RL96,BB02,NFG99,AF99} do not
suffer from this downside. In these
mechanisms, principals possess digital signature public and secret key pairs along with signed certificates binding
 authorizations to their public key. A principal makes requests by signing
statements using its secret key and  presenting one or more of its certificates. These certificates can
be validated and interpreted by any principal as long as it knows the public key of the certificate issuer.
Such a mechanism is used for authenticating HTTPS~\cite{DR08} servers on the Web.

Certificate-based mechanisms rely on a \emph{public-key infrastructure} (PKI) for distributing certificates
to various principals. Traditional x509 PKI~\cite{SFBHP08}, such as the one used on the Web, is centralized
with a hierarchical network of certification authorities responsible for issuing certificates. In light of the
downsides of centralization, several decentralized PKI~\cite{RL96,EFLRTY99,Z95,BFI99,BFL96} have
been proposed in the literature. A prominent model among these is the simple distributed security infrastructure
(SDSI)~\cite{RL96} of Rivest and Lampson. The SDSI model was subsequently merged with the 
simple public key infrastructure (SPKI) effort~\cite{EFLRTY99}, and the resulting model is commonly referred
to as SPKI/SDSI. In what follows, we briefly summarize some of the key ideas in SPKI/SDSI, while
referring the reader to~\cite{RL96,EFLRTY99} for a more comprehensive description.

\stitle{SPKI/SDSI}
This is a decentralized PKI based on the idea of \emph{local names}.
Each principal in this model is represented by a digital signature public key, and manages a local
name for referring to other principals. For instance, a principal Alice may use the name {\tt friend} to refer
to her friend Bob's public key and {\tt doctor} to refer to her family doctor Frank's public key. These bindings
are local to Alice, and other principals may chose to bind different names to these keys. However, another
principal, say Alice's TV, who names Alice's public key as {\tt Alice} may refer to Bob's
public key as {\tt Alice's friend}. Thus names in different namespaces can be linked using the {\tt 's} operator.
Local name to key bindings are represented by \emph{name-definition} certificates signed
by the issuing principal. Linked names are thus realized by certificate chains.
The model also supports \emph{authorization} certificates wherein an issuing
principal delegates permissions to another principal.

Access control policies in SPKI/SDSI specify a list of authorized principals using  local names
in the owner's namespace. A request is allowed by the policy if the requesting
principal is directly authorized by the policy or has a delegation (via authorization certificate)
from a directly authorized principal. For instance,  Alice's TV may have an access control
policy allowing the local name {\tt Alice's friend}, and therefore Bob (who has the name {\tt Alice's friend} in
the TV's namespace) and any principal delegated by Bob will have access.
Authorizing requests involves assembling a chain of certificates (from a repository of certificates)
that proves that the requesting principal satisfies the access control policy~\cite{CEEFMR99}. 
The responsibility of assembling the right certificate chain may be placed on the reference monitor
or the requester, and various deployments may differ in this choice. 
The key idea from SPKI/SDSI used in the Vanadium authorization model is that of delegating access
to principals by assigning them local names, and basing access control policies on these names.

\stitle{Caveats on delegation}
There are several mechanisms in the literature on restricting the scope of delegations.
These range from simple, coarse-grained mechanisms of adding a purpose and expiration time to
 delegation certificates, to complex, fine-grained mechanisms of extending
delegation certificates with S-expressions capturing application-specific permissions~\cite{EFLRTY99}, or
program code~\cite{BB02} defining how access must be proxied to the resource. 
Recently, Birgisson et al., proposed a mechanism for restricting delegations using \emph{caveats}~\cite{BPETVL14}, which
aims at striking a balance between simplicity and expressiveness.

Caveats are essentially predicates that restrict the context in which the delegated credential may be 
used. They are attached to the credential in a tamper-proof manner, each time the credential is delegated.
Caveats are of two types---\emph{first-party} and \emph{third-party}.
First-party caveats are predicates on the context in which a credential
may be used. 
For e.g., first-party caveats impose restrictions on the time of request (e.g.,
only between 6PM to 9PM), the permitted operation (e.g., only Read requests), 
the requester's IP address (e.g., the IP address must not be blacklisted), etc.
These restrictions are validated by the reference monitor in the
context of an incoming request, and the credential is considered invalid
if any caveat present on it is invalid.

Third-party caveats are restrictions wherein the burden of validation is pushed
to a third-party, i.e., neither the party that
wields the credential nor the party that is authorizing it. A credential with a
third-party caveat is considered
valid only when accompanied by a \emph{discharge} (proof of validity) issued by the
specific third party mentioned in the caveat. This discharge must be obtained
by the holder of the credential  before using the
credential as part of a request. A reference monitor making authorization decisions
simply checks that valid discharges are provided for all third-party caveats on 
the credential. 

A third-party caveat can be used for implementing revocation checks by having
the discharge service issue discharges only if the credential has not been revoked. The
discharge may be short-lived, and thus the holder of the credential would be obligated to 
periodically obtain fresh discharges from the service. Although this mechanism seems similar to the online certificate status protocol (OCSP)~\cite{MAMGA08}, the key difference is
that unlike an OCSP response, fetching a discharge  is the responsibility of the
principal making the request rather than the one authorizing it. 
Other examples of third-party caveats would be restrictions pointed at a social networking
service (e.g., discharged by checking membership in the ``work'' circle), or an 
auditing service (e.g., discharged by adding an entry to the audit log).  Discharges may themselves
carry first-party and third-party caveats thereby making the overall mechanism highly expressive.
For instance, a \emph{parental-control} caveat on  a credential handed to a kid  may initially point to a service
on dad's phone who in some cases may issue a discharge with a third-party caveat pointed at mom's phone.
While caveats were originally designed in the context of Macaroons~\cite{BPETVL14}, in this
work we use them to restrict the scope of delegation certificates in Vanadium.

\section{Vanadium Authorization Model}\label{sec:vanadium}
In this section, we describe the authorization model
of the \emph{Vanadium} framework~\cite{Vanadium}.
Vanadium is a set of tools, libraries, and services for developing secure
distributed applications that can run over a network of devices. At the core of
the framework is a remote procedure call (RPC) system that enables applications
to expose services over the network.  The framework offers an interface
definition language (IDL) for defining services, a federated naming system
for addressing them, and an API for discovering accessible services.
The authorization model is responsible for controlling access to RPC services, and 
ensuring that all RPCs are end-to-end encrypted, mutually authenticated, and
mutually authorized. 

\stitle{Preliminaries}
The model makes use of a digital signature scheme (e.g., ECDSA P-256).
In particular, we assume public and secret
 key pairs ($\pk, \sk$), and algorithms
$\sign$ and $\verifysig$ for signing messages and verifying signatures respectively.
$\sign(\sk, \msg)$ uses a secret key $\sk$ to produce a signature over a message $\msg$,
and $\verifysig(\pk, \msg, \sig)$ verifies a signature $\sig$ over a message $\msg$ using a
public key $\pk$. For any public and secret key pairs
$\pk, \sk$, $\forall \msg: \verifysig(\pk, \msg, \sign(\sk, \msg))$ holds.
We also assume a cryptographically secure hash function (e.g., SHA256), denoted
by $\hash$. For convenience, we assume that $\hash$ takes an arbitrary number of
arguments of arbitrary type, and internally encodes all arguments into a byte array using
some lossless encoding technique.

\subsection{Principals and Blessings}
A principal is any entity that can interact in the Vanadium framework.
Specifically, processes, applications, and services that include a
Vanadium runtime are all principals.  Each principal is associated
with a public and secret key pair, with the secret key never being shared
over the network. Each principal has one or more hierarchical human-readable
names associated with it called \emph{blessings}.
For instance, a television set (TV) owned by a user Alice\footnote{For ease of discussion, we refer to users and devices
as principals; strictly speaking, we are referring to processes controlled by them.}
 may have a blessing {\tt Alice{\sep}TV}.
Principals can have multiple blessings, and thus the same
TV may also have a blessing {\tt PopularCorp{\sep}TV123} from its manufacturer.

Principals are authenticated and authorized during requests based on the
blessing names bound to them. The public key of the principal does not
matter as long as the principal can prove that it has a blessing name
satisfying the other end's access control policy. We believe that this choice
makes it easier and more natural for users and system administrators to define
access control policies and inspect audit trails as they have to reason only
in terms of human-readable names.
%
\begin{figure}
\[\begin{array}{rll}
  \mydefhead{n \synteq}{\mbox{ordinary names not containing \sep}}{names}
  \mydefhead{\cert \synteq}{\certcon{n, \pk, \clist, \sig}}{certificates}
  \mydefhead{\bl \synteq}{\cert}{blessings}
  \mydefcase{\bl{\lsep}\cert}{}
  \mydefhead{\clist \synteq}{\emptylist}{list of caveats}
  \mydefcase{\clist{\lsep}\cav}{}
  \\
  \mydefhead{\cav \synteq}{\fpcav ~\mid~ \tpcav}{caveats}
  \mydefhead{\fpcav \synteq}{\timecav}{first-party caveats}
 \mydefcase{\targetcav}{}
 \mydefcase{\ldots}{}
  \mydefhead{\tpcav \synteq}{\tpccon{\nonce, \pk, \fpcav, \loc}}{third-party caveats}
  \mydefhead{\dis \synteq}{\discon{\clist, \sig}}{discharges}
\end{array}\]
\caption{Certificates, Blessings and Caveats}\label{fig:blessings}
\end{figure}
Concretely, a blessing is represented via a chain of certificates. The formal definition of
certificates and blessings is given in Figure~\ref{fig:blessings}. We use $\la \ldots \ra$ to
define tuples, colon as a binary operator for forming lists, and $\emptylist$ for the empty list.
$n$ ranges over ordinary names not containing $\sep$.  
Certificates, denoted by $\cert$, contain exactly four fields---a name, a public key, a (possibly empty)
list of caveats, and a digital signature. We
discuss the definition of caveats a bit later; for now they can be thought of 
as restrictions (e.g., expiration time) on the validity of the certificate. We use the dot notation to refer to
fields of a tuple, and thus $\cert.n$ is the name of a certificate $\cert$. Notice that our certificate
format is much simpler in contrast to x509 certificates~\cite{SFBHP08}. 

Blessings (denoted by $\bl$) are non-empty lists of certificates with each certificate
capturing a component of the blessing name. The list of certificates is meant to form a chain such
that signature of each certificate except the first one can be verified using the public key of the
previous certificate. The first certificate is self-signed, that is, its signature can be verified by its
own public key.  The public key listed in the final certificate
is the public key of the principal to which the blessing is bound. This public key
is denoted by $\fpubkey(\bl)$ for a blessing $\bl$. The name of a blessing
is obtained by concatenating all names appearing in the
blessing's certificate chain using $\sep$. This name is denoted by $\fname(\bl)$.

As an example, the blessing {\tt PopularCorp{\sep}TV123} is bound to the television's public key
$\pk_{TV}$ by a chain of two certificates---(1) a certificate with name {\tt PopularCorp}
and public key $\pk_{PopularCorp}$, signed by $\sk_{PopularCorp}$, and (2) a certificate
with name  {\tt TV123} and public key $\pk_{TV}$, signed by $\sk_{PopularCorp}$.
The validity of the certificate chain of a blessing $\bl$ is defined by the predicate
$\verifycerts(\bl)$.
\[
\begin{array}{l}
\verifycerts(\bl) \synteq \\
\quad \prcase \bl \prof\\
\quad \begin{array}{lll}
\mydefhead{\quad\phantom{\mid\ } \cert }{\prthen\verifysig(\cert.\pk, \hash(\cert.n, \cert.\pk, \cert.\clist), \cert.\sig)}{}
\mydefhead{\quad\mid\ \bl{\lsep}\cert }{\prthen\verifycerts(\bl) \land \verifysig(\fpubkey(\bl), \hash(\bl,\cert.n, \cert.\pk, \cert.\clist),\cert.\sig)}{}
\end{array}
\end{array}
\]
The signature in each certificate of the chain is not just over the certificate's contents
but also the chain leading up to the certificate. This means that each certificate
is cryptographically integrated into the blessing, and cannot be extracted and used
in another blessing. This property does not hold for SPKI/SDSI and
many other certificate-based systems, where certificates can be chained
together in arbitrary ways. As
a result, Vanadium does not face the \emph{certificate chain discovery} problem
\cite{CEEFMR99} that involves assembling the right chain of certificates (from a repository)
to prove that certain credentials are associated with a  public key.
A Vanadium blessing is a tightly bound certificate chain containing all the certificates required to
prove that a certain name is bound to a public key.

%
		
\subsection{Delegation and Caveats}
Blessings can be delegated from one principal to another by
extending them with additional names. For instance, Alice may extend her
blessing {\tt Alice} to her TV as {\tt Alice{\sep}TV}.
The TV may in turn extend this blessing to the Youtube application running on it
as {\tt Alice{\sep}TV{\sep}Youtube}.  Since principals possess authority
by virtue of their blessing names, delegating a blessing amounts to
delegation of authority. For instance, delegation of the blessing
{\tt Alice{\sep}TV} allows the TV to access all resources
protected by an ACL of the form {\tt Allow Alice{\sep}TV}.

Concretely, a delegation is carried out by the operation $\bless(\pk, \sk, \bl, n, \clist)$
which takes the public key $\pk_{d}$ of the delegate, the secret key $\sk$ of the delegator, the
blessing $\bl$ that must be extended, the name $n$ used for the extension, and the list
of caveats $\clist$ on the delegation. It extends the blessing's certificate chain
with a certificate containing name $n$, public key $\pk_{d}$, list of caveats $\clist$, and signed by the secret key $\sk$.
\[
\bless(\pk_{d}, \sk, \bl, n, \clist) \synteq \bl{\lsep}\certcon{n, \pk_{d}, \clist, \sign(\sk, \hash(\bl, n, \pk, \clist))}
\]
It is easy to see that if the blessing $\bl$ is valid, and the secret key $\sk$ corresponds
to the public key $\fpubkey(\bl)$, then the blessing $\bless(\pk_{d}, \sk, \bl, n, \clist)$ is valid.
Each blessing delegation
involves the blesser choosing an extension  (e.g., {\tt TV}) for the blessing. 
The extension may itself
have multiple components (e.g., {\tt home{\sep}bedroom{\sep}TV}), and a blesser may
choose the same or a different extension across multiple delegations. The role of the
extension is to namespace the delegation, similar to \emph{local names} in SDSI~\cite{RL96}.

The role of the caveat list
($\clist$) supplied to the $\bless$ operation is to restrict the conditions under
which the resulting blessing can be used.
For example, Alice can bless her TV as {\tt Alice{\sep}TV} but
with the caveats that the blessing can be used only between 
{\tt 6PM} and {\tt 9PM}, and only to make requests to her video service
(and not her bank!).
Thus,
the resulting blessing makes the assertion:
\begin{quote}
The name {\tt Alice:TV} is bound to $\pk_{TV}$\\
  \emph{as long as} the TIME is between {\tt 6PM} and {\tt 9PM}, and \\
  \emph{as long as} target of the request matches {\tt SomeCorp{\sep}VideoService} \\
\end{quote}
When the TV uses this blessing to make a request to the video service, the
service will grant the request only if the current time is within the permitted
range, and its own blessing name matches {\tt SomeCorp{\sep}VideoService}.

The caveats in the above example are all first-party caveats as they are
validated by the reference monitor at the target service when the blessing is
presented during a request.
As discussed in Section~\ref{sec:ideas} and~\cite{BPETVL14}, first-party
caveats are validated in the \emph{context} of an incoming request based on
 information such as the time of request,
the blessing presented, the method invoked, etc. We use $\con$ to denote request
contexts, and assume a function $\validatefpcav(\fpcav, \con)$ that checks if
a first-party caveat $\fpcav$ is valid in the context $\con$. 
%

While the Vanadium framework implements $\validatefpcav$ on some standard first-party
caveats (e.g., expiry, method and peer restriction), services may also define their own first-party caveats.
For instance, a video streaming service may define a ``PG-13" caveat so that blessings carrying
this caveat would be authorized to stream only PG-13 movies.
%

\subsection{Third-Party Caveats}
Vanadium blessings may also carry third-party caveats~\cite{BPETVL14} wherein
the burden of validating the caveat is pushed to a third-party.
For example, Alice can bless
Bob as {\tt Alice{\sep}Houseguest{\sep}Bob} but with the caveat that the blessing is valid
only if Bob is within $20$ feet of Alice as determined by a proximity service running on Alice's phone (third-party). Before making a request with this blessing, Bob must obtain a
\emph{discharge} (proof) from the proximity service on Alice's phone. This discharge must be sent along with the blessing in a request. Thus, the blessing makes the assertion:
\begin{quote}
The name  {\tt Alice{\sep}Houseguest{\sep}Bob} is bound to $\pk_{Bob}$\\
\emph{as long as} a proximity service on Alice's phone issues a discharge after validating
that Bob is ``within $20$ feet" of it.
\end{quote}
The structure of a third-party caveat and discharge is defined in Figure~\ref{fig:blessings}.
Every third-party caveat includes a nonce (for uniqueness), a public key controlled by the
third-party service, a first-party caveat specifying the
check that must carried out  before issuing the discharge, and the location of the third-party
service. A discharge contains a signature and possibly additional caveats.
It is considered valid if its signature can be verified by the
public key listed in the third-party caveat, and any additional caveats listed on it are valid.

The operation $\mintdischarge(\sk_{tp}, \tpcav, \clist, \con)$ uses a secret key $\sk_{tp}$
(owned by the third-party) to produce a discharge for a third-party caveat $\tpcav$
if the check specified in $\tpcav$ is valid in the context $\con$. The returned discharge
contains the provided list of caveats $\clist$.
\[
\begin{array}{l}
\mintdischarge(\sk_{tp}, \tpcav, \clist, \con) \synteq \\
\quad \prcase \validatefpcav(\tpcav.\fpcav, \con) \prof\\
\quad \begin{array}{lll}
\mydefhead{\quad \phantom{\mid\ } \prtrue}{\prthen \prerror}{}
\mydefhead{\quad \mid\ \prfalse}{\prthen \discon{\clist, \sign(\sk_{tp}, \hash(\tpcav, \clist))}}{}
\end{array}
\end{array}
\]
The purpose of the additional caveats on the discharge is to limit its validity. 
For instance, in the proximity caveat example, Alice's phone may issue a discharge
that expires in $5$ minutes thereby requiring ${\tt Bob}$ to fetch a new discharge
periodically. This ensures that Bob cannot cheat by fetching a discharge when near
Alice's phone, and then using it later from a different location. 

\subsection{Blessing Roots}
Since blessing names are the basis of authorization, it is important
for them to be unforgeable.  Simply verifying the signatures in a certificate chain
does not protect against forgery because the first certificate in the chain is
self-signed, and thus can be constructed by any principal.
For instance, an attacker with public and secret keys $\pk_{a}, \sk_{a}$ can bind the
name {\tt Alice} to her public key using the blessing 
$\certcon{{\tt Alice}, \pk_a, \emptylist, \sign(\sk_a, \hash({\tt Alice}, \pk_a, \emptylist))}$.
She may then extend this blessing with any extension of her choosing using the $\bless$
operation.

In order to defend against such forgery, all Vanadium principals have a set of 
\emph{recognized roots}. The \emph{root} of a blessing is the public key and name
of the first certificate in the blessing's certificate chain. A blessing is recognized by a principal only if
the blessing's root is recognized. For instance, all of Alice's devices may recognize her
public key $\pk_{Alice}$ and name ${\tt Alice}$ as a root. Any blessing prefixed with {\tt Alice}
would be recognized only if it has $\pk_{Alice}$ as the root public key. Similarly all
devices manufactured by PopularCorp may 
recognize the public key  $\pk_{PopularCorp}$ and name {\tt PopularCorp} 
as a root. Devices from another manufacturer may not recognize this root, and 
would simply discard  blessings from PopularCorp.

We use the term \emph{identity provider} for a principal that acts as a blessing root and hands
blessings to other principals. For instance, both
Alice and PopularCorp would be considered as identity providers by Alice's TV. Any principal can
become an identity provider. In general, we anticipate well-known companies, schools, and public
agencies to become identity providers, and different principals may recognize different subsets of
these.

\subsection{Authentication and Authorization}
Client and servers in a remote procedure call (RPC) authenticate each other by
presenting blessings bound to their public keys. The Vanadium authentication
protocol~\cite{vanadium-auth-protocol} is based on the well-known SIGMA-I protocol~\cite{CK02}.
It ensures that each end learns the other end's blessing, and is convinced that the other end possesses the corresponding secret key. At the end of the protocol, an
encrypted channel (based on a shared key) is established between the client
and server for further communication. Since the protocol is fairly standard,
we do not discuss it here and refer the reader to~\cite{vanadium-auth-protocol} 
for a detailed description.

Once authentication completes, each end checks that the other end's blessing
is authorized for the RPC. Authorization is mutual. 
For example, Alice (client) may invoke a
method on her TV (server) only if the TV presents a blessing matching
{\tt Alice{\sep}TV}, and the TV may authorize Alice's
request only if she presents a blessing prefixed with {\tt Alice}.
Authorization checks involve two key steps---(1) validating the blessing presented by the other
end, followed by (2) matching the blessing name against an access control policy.

A blessing is always validated in the context of a given request. Blessing validation
involves ---(1) verifying the signatures on the blessing's certificate
chain, (2) verifying that the blessing root is recognized,
and (3) validating all caveats on the blessing
in the context of the request. 

$\validatebl(\bl, \roots, \con)$, defined below, 
determines if a blessing $\bl$ is valid for a given request context $\con$ and
a set of recognized roots $\roots$. The context is assumed to contain all parameters
of the request, including any discharges sent by the other end. Specifically, 
$\fdis(\con)$ denotes the discharges contained in the context $\con$,
$\froot(\bl)$ is the root of the blessing $\bl$ and $\fcavs(\bl)$ is the set of all caveats appearing
on certificates of the blessing $\bl$. A first-party caveat is validated by invoking the function $\validatefpcav$, 
and a third-party caveat is validated by finding a matching discharge in the request context, i.e., 
a discharge whose  signature can be verified using the public key specified in the third-party
caveat. Additionally, any caveats listed on the discharge are also (recursively) validated.
\[
\begin{array}{l}
\validatebl(\bl, \roots, \con) \synteq \\
\quad \verifycerts(\bl) \land \isrec(\bl, \roots) \land \validatecavs(\fcavs(\bl), \con) \\
\\
\isrec(\bl, \roots)  \synteq  \exists r \in \roots: \froot(\bl) = r \\
\\
\validatecavs(\clist, \con)  \synteq \\
\quad \forall \cav \in \clist: \\
\quad\quad \prcase \cav \prof \\
\quad\quad \begin{array}{lll}
\mydefhead{\quad\phantom{\mid\ } \fpcav}{\prthen\validatefpcav(\fpcav, \con)}{}
\mydefhead{\quad \mid\ \tpcav}{\prthen \exists \dis \in \fdis(\con):}{}
\mydefhead{}{\quad~ \verifysig(\tpcav.\pk, \hash(\tpcav, \dis.\clist), \dis.\sig)) \land \validatecavs(\dis.\clist, \con)}{}
\end{array}
\end{array}
\]

Once the remote end's blessing is validated, the name of the blessing
 is checked against an access control policy. Access is granted only if the check
is successful. We discuss the structure of these policies next.

\subsection{Access Control Policies}
The syntax and semantics of access control policies in Vanadium has been defined rigorously
in~\cite{ABPSST15}. We give a brief overview of the design in this subsection. Policies in Vanadium are specified using access control lists (ACLs) that 
resolve to the set of  permitted blessing names.
In order to allow policies to be short and simple, Vanadium allows ACLs
to indirect through groups. For example, Alice may create a group ${\tt \grp{AliceFriends}}$
containing the blessing names of all her friends and add it to all relevant ACLs.
This saves her from enumerating the list
of blessing names of family members within each ACL, and also provides
a central place to manage multiple policies. Furthermore, group definitions may be nested. 
For instance, the definition of the group ${\tt \grp{AliceFriends}}$ may contain the group
${\tt \grp{DaveFriends}}$ containing the friends of Alice's flatmate Dave. 
Typically, the definition of a group would be held at a remote server, which
would be contacted during ACL resolution. The ACL resolver may cache information
about groups in order to combat unreliable network connectivity and avoid expensive
network roundtrips.

\begin{figure} 
\[\begin{array}{rll}
  \mydefhead{\bn \synteq}{n}{blessing names}
  \mydefcase{n{\sep}\bn}{}
  \mydefhead{\bp \synteq}{n}{blessing patterns}
  \mydefcase{g}{}
  \mydefcase{n{\sep}\bp}{}
  \mydefcase{g{\sep}\bp}{}
  \mydefhead{\lbp \synteq}{\emptylist}{list of blessing patterns}
  \mydefcase{\lbp{\lsep}\bp}{}
  \mydefhead{\acl \synteq}{\allow{\lbp}~~\deny{\lbp}}{ACL}
\end{array}\]
\caption{Access control policies}\label{fig:authpolicies}
\end{figure}
ACLs may also contain blessing names where one of the components is a group name.
Such names are called \emph{blessing patterns}, and are meant to capture a derived
set of blessing names. For example, the pattern ${\tt \grp{AliceFriends}{\sep}Phone}$
defines the set of blessing names of phones of Alice's friends. In particular,
if the group ${\tt \grp{AliceFriends}}$ contains the blessing name {\tt Bob}, then
the pattern ${\tt \grp{AliceFriends}{\sep}Phone}$ would be matched by the blessing
name {\tt Bob{\sep}Phone}.

\stitle{Definitions} 
The formal definition of blessing patterns and ACLs is given in  Figure~\ref{fig:authpolicies}.
As before, $n$ ranges over ordinary names not including {\sep}. $g$ ranges over
group names (i.e., name with the subscript ``{\tt G}''), and $\bn$ ranges over blessing names.
A blessing pattern is a non-empty sequence of ordinary names or group names separated
by {\sep}. An ACL is a pair of $\allows$ and $\denys$ clauses, each containing a
list of blessing patterns.\footnote{The model described in~\cite{ABPSST15} is more general
and allows multiple $\allows$ and $\denys$ clauses in ACLs.} $\denys$ clauses make it convenient to encode blacklists.
For instance, the ACL $\allow{\tt \grp{AliceDevices}}~\deny{\tt \grp{AliceWorkDevices}}$
allows access to all of Alice's devices except her work devices.

Group definitions are of the form $g \eqdef \lbp$,  and thus equate a group name with a list of
blessing patterns. For example, the group ${\tt \grp{AliceFriends}}$ may have a definition of the form
\[
{\tt \grp{AliceFriends}} \eqdef {\tt Bob}, {\tt Carol}, {\tt \grp{DaveFriends}}
\] 
A given blessing name satisfies an ACL if there is at least one blessing pattern in the $\allows$ clause
that is matched by the blessing name, and no blessing pattern in the $\denys$ clause is
matched by the blessing name. For example, when {\tt Bob} is in the group ${\tt \grp{AliceFriends}}$, the
ACL ${\allow{\tt \grp{AliceFriends}}}$ will permit access with the blessing name {\tt Bob} but the ACL 
${\allow{\tt \grp{AliceFriends}}~\deny{{\tt Bob}}}$ will deny it. The default is to deny access, 
so for example the ACL $\allow{{\tt Bob}}$ will deny access to {\tt Carol}. 

The semantics of ACL checks makes use of the prefix relation on blessing names.
Given two blessing names $\bn_1$ and $\bn_2$, we write
$\bn_1 \bprefix \bn_2$ 	if the sequence of names in $\bn_1$ (separated by $\sep$) is 
a prefix of the sequence of names in $\bn_2$, for e.g., ${\tt Alice} \bprefix {\tt Alice{\sep}TV}$
holds but  ${\tt Ali} \bprefix {\tt Alice{\sep}TV}$ does not.

In order to formalize the ACL checking procedure, we first define a function 
$\fmeaning$ that maps a blessing pattern to a set of blessing names. It is parametric
on a semantics of group names, which is a function $\rho$ that maps a group
name to a set of members of the group. We discuss how $\rho$ is obtained later.
\[
\begin{array}{l}
\fmeaning(\bp) \synteq \\
\quad \prcase \bp \prof\\
\quad \begin{array}{lll}
\mydefhead{\quad \phantom{\mid\ } n}{\prthen \{ n \}}{}
\mydefhead{\quad \mid\ g}{\prthen \rho(g)}{}
\mydefhead{\quad \mid\ n {\sep} \bp}{\prthen \{ n {\sep} s \mid s \in \fmeaning(\bp) \}}{}
\mydefhead{\quad \mid\ g {\sep} \bp}{\prthen \{ s {\sep} s' \mid s \in \fmeaning(g), s' \in \fmeaning(\bp)\}}{}
\end{array}
\end{array}
\]
$\fmeaning$ can be naturally extended to a list of blessing patterns by defining it as the
union of the sets obtained by applying $\fmeaning$ to each element of the list, with
$\fmeaning(\emptylist)$ defined as $\emptyset$. 
Using the function $\fmeaning$, we define the function $\isauthorized(\bn, \acl)$ that decides
whether a given blessing name $\bn$ (seen during a request) satisfies an ACL $\acl$.
\[
\begin{array}{l}
\isauthorized(\bn, \acl) \synteq \\
\quad \prcase \acl \prof\\
\quad \begin{array}{lll}
\mydefhead{\quad \phantom{\mid\ } \allow{\lbp_A}~~\deny{\lbp_D}}{\prthen (\exists \bn' \in \fmeaning(\lbp_A).\/ \bn'  \bprefix  \bn)}{}
\mydefhead{}{\land (\lnot \exists \bn' \in \fmeaning(\lbp_D).\/ \bn'  \bprefix  \bn)}{}
\end{array}
\end{array}
\]
The function checks that the blessing name $\bn$ matches an allowed blessing pattern and does
not match any denied blessing pattern. Matching is defined using prefixes (instead of exact
equality) for both allowed and denied clauses; the reasons however are different. 

For $\allows$ clauses, the use of the relation $\bprefix$  is a matter of convenience. We believe
that often when a service grant access to a principal (e.g., with blessing name {\tt Alice}) 
it may be fine if the access is exercised by delegates of the principal (e.g., with blessing
name {\tt Alice/Phone}). Thus a pattern in an $\allows$ clause is considered matched as long
as some prefix of the provided blessing name matches it. 
Alternatively, services that want to force exact matching for allowed patterns---perhaps
to prevent the granted access from naturally flowing over to delegates---may use
the special reserved name $\bdollar$ at the end of the pattern. For e.g., the pattern
$\allow{\tt Alice{\sep}\bdollar}$ is matched only by the blessing name ${\tt Alice}$.

For $\denys$ clauses, ensuring that no prefix of the blessing name matches a denied pattern
is crucial for security. A principal with blessing name $\bn$ can always extend it 
(using the $\bless$ operation) and bind it to itself. Thus, from a security perspective it is important that if $\bn$ is denied access, then all extensions of $\bn$ are also denied access.

\stitle{Semantics of groups ($\rho$)}
We now discuss how the map $\rho$ from group names to members of the group is defined.
In Vanadium, groups may also be distributed, in that, different group definitions may be held
at different servers. Given this, defining the map $\rho$ becomes complicated for several
reasons. Firstly, due to network partitions some group servers may be unreachable during 
ACL checking and thus their definitions may be unavailable. The definition of a group may 
depend on other groups, and there may be no  overseeing authority ensuring absence of 
dependency cycles. Finally, group server checks need to be secure and private. For instance, 
group servers under different administrative domains may be unwilling to reveal their complete 
membership lists to each other, and may offer only an interface for membership lookups. We 
refer the reader to~\cite{ABPSST15} for a complete treatment of how these issues are tackled,
and explain only the key ideas below.

When group servers are unreachable during ACL checking, we conservatively approximate the
definition of $\rho$ for those groups. The approximation depends on whether the group is being 
resolved in the context of an $\allows$ clause or a $\denys$ clause. While matching an allowed 
pattern, unreachable groups are under-approximated by the empty set. On the other hand, while 
matching a denied pattern, unreachable groups are over-approximated by the set of
all blessings. Thus effectively we define two maps $\rhounder$ and $\rhoover$---the map $\rhounder$  is used while defining $\fmeaning$ for allowed patterns, and the map $\rhoover$ is used while defining  $\fmeaning$ for denied patterns. The maps $\rhounder$ and $\rhoover$ coincide when all group 
definitions are available.

 ${\rhounder}$ and $\rhoover$ can be constructed by considering the list of available group definitions 
 as a set of productions inducing a formal language. Ordinary names and ${\sep}$
are terminals, and group names are non-terminals. For instance, the group definition
\[
g_1 \eqdef {\tt Alice}{\sep}{\tt Phone},\; g_2 {\sep} {\tt Phone}
\] 
can be viewed as two productions
\[
\begin{array}{rcl}
g_1 & \rightarrow & {\tt Alice}{\sep}{\tt Phone}\\
g_1 & \rightarrow & g_2{\sep}{\tt Phone}
\end{array}
\]
For $\rhounder$, no production is associated with a group name whose definition is unavailable.
On the other hand for $\rhoover$, such group names are associated with productions inducing the set 
of all blessings. Once the set of productions are defined, for any group name $g$, $\rhounder(g)$
and $\rhoover(g)$ are defined as the set of blessing names generated from $g$ 
by the corresponding set of productions. 

While constructing $\rhounder$ and $\rhoover$ in the aforesaid manner is infeasible in practice
as it requires knowledge of all the group definitions, the key observation here is that checking group membership can be reduced to checking membership in an induced formal language. In~\cite{ABPSST15}, 
this observation and techniques from top-down parsing are
used  to develop a distributed algorithm for checking whether a blessing name belongs to a group.

\subsection{Life of an RPC}
We now explain how the various parts of the authorization model 
come together during the course of an RPC.
Consider Alice's house guest Bob who wants to  invoke a method on Alice's TV.
Suppose Bob has the blessing $\bl_{Bob}$ with name 
{\tt Alice{\sep}Houseguest{\sep}Bob}, and the TV has the blessing $\bl_{TV}$
with name {\tt Alice{\sep}TV}. Additionally, suppose
that Bob's blessing has a third-party caveat to a proximity service running on
Alice's phone. Bob has the access control policy $\allow{\tt Alice{\sep}TV}$
that allows only Alice's TV, and the TV has the access control
policy $\allow{\tt {Alice}{\lsep}{Alice{\sep}Houseguest}}$ that allows only Alice and
her house guests.
In what follows, we describe the steps involved in the RPC from Bob's phone
to Alice's TV. We focus on the authentication and access-control
aspects, and do not discuss how various network connections are
established.
\begin{itemize}
\item Bob uses the Vanadium authentication protocol to initiate a connection
to Alice's TV. 
\item As part of the exchange, the TV first sends its blessing $\bl_{TV}$
to Bob. Bob invokes $\validatebl(\bl_{TV}, \roots_{Bob}, \con_{Bob})$ 
to verify that the blessing $\bl_{TV}$ is valid for the current context from Bob's
perspective and the set of blessing roots
recognized by Bob ($\roots_{Bob})$. If this step succeeds, then Bob verifies that
 the name of the blessing (${\tt Alice{\sep}TV}$) satisfies his
 access control policy ($\allow{\tt Alice{\sep}{TV}}$).
 The connection is aborted if any of these checks fail.
\item After authorizing the TV's blessing, Bob selects his blessing $\bl_{Bob}$ 
(from Alice) to present to the TV. Since the blessing carries a third-party caveat,
Bob first connects to the third-party service listed on the caveat
to obtain a discharge. The service performs the necessary checks, and if those succeed,
it issues a discharge to Bob. Bob (recursively) performs
the above procedure for any third-party caveats on the discharge, and once all discharges have been obtained, he presents all of them with the blessing $\bl_{Bob}$ to the TV.
\item The TV invokes $\validatebl(\bl_{Bob}, \roots_{TV}, \con_{TV})$ to verify that the
blessing $\bl_{Bob}$ is valid for the current context from the TV's perpspective ($\con_{TV}$)
and the set of blessing roots recognized by the TV ($\roots_{TV}$).
If this step succeeds, the TV verifies that the name of the blessing
(${\tt Alice{\sep}Houseguest{\sep}Bob}$) 
satisfies its access control policy ($\allow{\tt {Alice}{\lsep}{Alice{\sep}Houseguest}}$).
The connection is aborted if any of these checks fail.
\item After authorization succeeds at the TV's end, the protocol is complete and
an encrypted channel is established between Bob and the TV. Application data pertaining
to the RPC is then exchanged on this channel.
\end{itemize}

\subsection{Practical Considerations}\label{sec:practicalities}
We now discuss some considerations involved in deploying
the authorization model in practice.

\stitle{Managing blessings}
Authorization in Vanadium is based on blessings. A principal may acquire
multiple blessings over time, each providing access to some set of services
under some contextual restrictions. Consequently, managing these blessings may become
quite onerous. The first problem is storing all these blessings while keeping
track of the meta-data about where they were obtained from and under  what constraints.
Another problem is selecting which blessing to present when authenticating
to a peer. While presenting all blessings and letting the peer choose the relevant one
is convenient, it has the downside of leaking sensitive information, for e.g., a blessing may
reveal that Bob is a house guest of Alice.
Instead, Vanadium provides a means to selectively share blessings with appropriate 
peers. Blessing are stored by Vanadium principals using a mechanism similar to cookie
jars in Web browsers. All blessings are stored with a blessing pattern identifying the
peer to whom they should be presented. This pattern may be set based on information
provided by the  blessing granter.  For example, Bob can add the blessing 
{\tt Alice{\sep}Houseguest{\sep}Bob} with the peer pattern {\tt Alice}. Thus, Bob will present this
blessing only when communicating with services that have a blessing name matching this pattern.
Any other service that Bob communicates with will not know that he has this blessing from Alice.

\stitle{Blessing vs. Adding to an ACL}
The careful reader may have noticed that Vanadium offers two mechanisms for granting
access to a resource.
For instance, consider Alice's TV with an ACL $\allow{\tt Alice}$ which means all principals 
with the blessing name {\tt Alice} or an extension of it have access. Alice can grant
access to her TV to another principal by either extending her {\tt Alice} blessing 
to the other principal or by adding the other principal's blessing name to the ACL.
The question then is how does one decide which method is appropriate in a given use-case. 
Granting a blessing is akin to handing out a capability, and thus in a way this question is that of deciding between granting a capability versus modifying an ACL. We recommend the following approach for making the choice.
%
%
%

 The constraints on the access being delegated must be taken into consideration. If the access is meant to be
 long-lived and unconstrained then modifying the ACL is preferable as it allows the service administrator to audit
 and revoke access at any time. For instance, Alice may share access to her TV with her flatmate Dave by adding
 the pattern {\tt Dave} to the TV's ACL. Later when Alice moves out, she can revoke Dave's access by removing
 this pattern. On the other hand, when the access is constrained then blessing with caveats is a more appropriate choice.
 For instance, Alice may delegate temporary access to her TV to her house guest Bob by blessing him under a short-lived
 time caveat. 
 
The choice of how access is granted also affects the subsequent auditability of the delegated access. For instance, when Dave accesses Alice' TV he would use his own blessing (assuming Alice's TV  trusts Dave as a blessing root) and his access would be recorded as {\tt Dave}. However, when Bob accesses
 the TV he would use his blessing from Alice, and thus his access would be recorded as {\tt Alice{\sep}Houseguest{\sep}Bob}. Finally, we note that the option to change an ACL may not always be available. When Bob wants
 to grant access to Alice's TV to his friend Carol,  extending his blessing to Carol may be his only option as he may not have the authority to modify the TV's ACL.

\stitle{Revocation}
We now discuss mechanisms for revoking access. Revocation is easy when access is granted by adding to an ACL or a group, as it amounts to simply removing the added entry. It is more challenging when access is granted via blessing. One approach is to always constrain blessings with short-lived time caveats, thereby invalidating
them automatically after a certain time. Principals would then have to periodically reach out to their blessing granters
for a fresh blessing. This idea is similar to ``reconfirmation" in SDSI~\cite{RL96}. Using a
 third-party caveat pointed at a revocation service offers a more systematic way of realizing this idea. The revocation service would issue a short-lived discharge for the caveat only if the blessing has not been revoked.

While such third-party caveats elegantly encode revocation restrictions, they suffer from the downside of requiring blessing holders to periodically connect to a revocation service. Specifically, they introduce a trade-off between how swiftly a blessing may be revoked, and how long things may operate when disconnected. This trade-off may be ameliorated to some extent if devices can recognize when they are offline and use different revocation timeouts in that case. Furthermore, in the home setting, a discharge service may be run on a WiFi access point, thereby requiring devices to maintain connectivity only to the local WiFi network.

\section{Application: Physical Lock}\label{sec:lock}
In this section, we explain how the Vanadium authorization model 
may be applied to a physical lock. The application highlights the
flexibility and decentralization aspects of the model. A  network controlled
lock is a common device found in many modern homes today.
It allows a user to lock and unlock a door from their phone, and share
access to it with visitors. Today, there are a number of manufacturers
building locks for homes, garages, factory floors etc.

The authorization model for most existing products involves a global service
service in the cloud, often controlled by the lock manufacturer, that is an
authority on all credentials used to access the lock. 
Typically, the service must be accessed during setup and whenever access is
delegated. As discussed in Section~\ref{sec:desired}, this is undesirable as
communicating with global services requires internet access, which may not be
perfectly reliable. It can be quite frustrating for a user to be unable to
share access to a lock due to lack of internet connectivity at the time sharing
is initiated. Furthermore, compromising the manufacturer owned service may
allow attacker to unlock all locks managed by the manufacturer.  In what
follows, we present an authorization model for locks that is fully
decentralized, and does not depend on access to an external service or identity
provider.

\stitle{Overview} The key idea is to have the lock be its own identity provider. 
When the lock is set up by its owner, it creates a self-signed blessing for itself,
and extends this blessing to the owner. The blessing granted to the owner is effectively the
\emph{key} to the lock. All subsequent access to the lock is  restricted to clients that
can wield this blessing or extensions of it. Delegation of access is simply carried out
by extending the \emph{key} blessing.

\subsection{Authorization Details}
Consider a user Alice who just bought a brand new lock for the front door
of her house. We walk through the steps of setting up identity and access
control for the lock. We assume that Alice is interacting with the lock
using another device, say her phone.

\stitle{Claiming a new lock}  We assume that an out-of-box lock device comes with a 
pre-installed public and secret key pair, and a blessing from its manufacturer of
the form {\tt <manufacturer>{\sep}<serial no>}. The first step in setting up the lock is
for Alice to name the lock and obtain a blessing for subsequently interacting with it.
This is accomplished by invoking the {\tt Claim} method on the lock that
returns a blessing bound to the invoker's (Alice's) public key.

The invocation is through a Vanadium remote procedure call.
The invoker authorizes the lock by verifying that it presents a
blessing from the manufacturer with the expected serial number. An unconfigured
lock authorizes any principal to invoke the {\tt Claim} method on it, after
which it considers the setup process complete and no longer allows invocation
of that method.


After the {\tt Claim} invocation is authorized, the lock creates a self-signed
blessing with a name provided by the caller.  This blessing is presented by the
lock to authenticate to clients during all subsequent invocations. The lock
then acts as an identity provider and extends this blessing to the invoker's
public key (learned during authentication).  This granted blessing is called
the \emph{key} blessing of the lock.   The invoker saves this blessing for
subsequent interactions with the lock and also recognizes its root as an
identity provider. For instance, Alice may claim the lock on her front door
with the name {\tt AliceFrontDoor}. The lock will subsequently authenticate to
others as {\tt AliceFrontDoor}, and would grant the blessing {\tt
AliceFrontDoor{\sep}Key} to Alice (here {\tt Key} is the blessing extension
used by the lock).

\stitle{Locking and unlocking}
Once a lock has been claimed, the {\tt Claim} method is disabled and the lock instead
exposes the {\tt Lock} and {\tt Unlock} methods. The methods are protected by an ACL
that allows access only to clients that wield a blessing from the lock's identity provider.
In the above example, the ACL would be $\allow{\tt AliceFrontDoor}$,
which would be
matched by the blessing {\tt AliceFrontDoor{\sep}Key}.

\stitle{Delegating access}
Any extension of the key blessing also matches the ACL for the lock's methods,
and thus access to the lock can be delegated by extending this blessing.  As
usual, caveats can be added to the extension to restrict its scope. For
instance, Alice can extend her blessing {\tt AliceFrontDoor{\sep}Key} to her
house cleaner as {\tt AliceFrontDoor{\sep}Key{\sep}Cleaner} under a time caveat
that is valid only on Mondays between 8AM to 10AM.

\stitle{Auditing access}
The lock can keep track of the blessings used to access it, even ones that have
invalid caveats. Thus, Alice can inspect the log on the lock for auditing
access attempts made by her house cleaner. In particular, Alice can  detect if
the cleaner tried to access the lock outside of the agreed upon time (8AM to
10AM on Mondays) or if there is access by someone who has received a delegated
blessing from the cleaner.

\subsection{Discussion}
We highlight three distinguishing aspects of the authorization model presented in this
section.

\stitle{Decentralized} Each lock is an authority on the secrets and
credentials that can be used to access it. No external entity, including the lock
manufacturer, can mint credentials to access a claimed lock device. The 
credentials for accessing one lock are completely independent from those for accessing
another. Thus, attackers have no single point of attack to unlock multiple instances.

\stitle{No internet connectivity required} The authorization model does not
require the lock or the device interacting with it to have internet access at
any point, including during setup.  The model does not rely on any third-party
service or identity provider.

\stitle{Audited} The lock can keep track of when it was accessed, by whom
(represented by the blessings). Since blessings inherently capture a delegation
trail, the access log also conveys how the invoker obtained access.

Having highlighted the above advantages, we note that decentralization
comes at a cost. For instance, the lack of an authoritative source in the cloud
makes it hard to recover from loss or theft of blessings and secret keys.
Furthermore, users are responsible for managing multiple blessings and
keeping track of various delegations they make. We believe that carefully
designed user-interfaces and appropriate reset modes for the lock device can
help address some of these concerns.

\section{Conclusion}\label{sec:conclusion}
This tutorial presents the authorization model of the
Vanadium framework~\cite{Vanadium}. In this model, each
principal has a digital signature public and secret key pair,
and a set of hierarchical human-readable names
bound to its public key via certificate chains
called \emph{blessings}. All authorizations associated
with a principal are based on its blessing names. In particular, a principal
makes a request to a service by presenting one of its blessings
(using the Vanadium authentication protocol~\cite{vanadium-auth-protocol}),
and the request is authorized if the blessing name 
satisfies the service's access control policy.

A notable feature of the model is its support for decentralization
and fine-grained delegation. 
The model does not require the existence of special identity providers
 that are trusted by all  principals by default.
Instead, any principal may choose to become an identity provider and create blessings
for itself and other principals. Each principal has a choice over which other principals
it recognizes as identity providers. Only blessings from recognized
identity providers are considered valid. In practice, we anticipate that there will be 
a small set of large-scale identity providers that most people will commonly use in
 interactions with the wider world. 

Principals can delegate access by extending one or more of their blessings to other
principals. The scope of delegations can be constrained very finely by adding
caveats~\cite{BPETVL14} to the delegated blessing. In particular, blessings support
third-party caveats which allow
predicating delegations on consent by specific third-parties. Such caveats elegantly
support revocation, proximity-based restrictions, and audit requirements.
%
%

Access control policies in Vanadium may indirect through groups whose definition may
be distributed across multiple servers, possibly under different administrative
domains. While several access control mechanisms support groups, a distinguishing
aspect of our design is using groups to construct compound names. For instance,
the pattern ${\tt \grp{AliceFriends}{\sep}Phone}$, with ${\tt \grp{AliceFriends}}$ being
a group of blessing names of Alice's friends, defines the set of blessing names of phones
of Alice's friends. Such compound names coupled with negative clauses in ACLs make
the problem of checking group membership fairly complex, especially when the group
server may be unreachable. The Vanadium authorization model mitigates some of the difficulties by making simplifying choices. For instance, group definitions cannot contain negative clauses, and ACLs cannot be reused for defining groups or other ACLs. 
This tutorial provides a brief overview of our solution, with a more comprehensive
description available in~\cite{ABPSST15}.

\stitle{Future directions} 
There are several future directions for this line of work. The first and perhaps the most
important direction is on making the model and its primitives easily usable by end users.
This is paramount to the adoption of the model. Mechanisms for making the model
more usable may include designing intuitive user-interfaces for visualizing, granting
and revoking blessings, and conventions on blessing names that help write intelligible
access-control policies.

Another direction is that of enabling \emph{mutual} privacy in the Vanadium authentication
protocol. Currently, the protocol involves the server presenting its blessing before the client.
While this is beneficial to the client, as it may choose to not reveal its blessing after seeing
 the server's blessing, it is disadvantageous to the server. The server's blessing is
effectively revealed to anyone who connects to it, including active network attackers.
In fact, this lack of mutual privacy is common to many other mutual authentication protocols 
(such as TLS~\cite{DR08}, SIGMA-I~\cite{CK02}) wherein one of the parties must reveal its
identity first. In the scenarios considered in  this work, the participants may be personal
end-user devices neither of which is inclined to reveal its identity before learning the identity of its
peer. In ongoing work~\cite{WTSB16}, we are designing a private mutual authentication protocol
that allows each end to learn its peer's blessing only if it satisfies the
peer's authorization policy.
 
Finally, one may consider designing mechanisms for securely leveraging
external cloud-based services when internet access is available. For instance, a Cloud-based
service may be used as a transparent proxy for RPCs, as a revocation and auditing
service for blessing delegations, or as a readonly backup for data. In all cases, the goal would
be to leverage external Cloud-based services for various tasks while granting them the minimal
authority necessary for the task.
%

\subsubsection*{Acknowledgments} This work is a result of a joint effort by several members of the
Vanadium team at Google. We would like to thank Mart{\'i}n Abadi, Jungho Ahn, 
Mike Burrows, Ryan Brown, Bogdan Caprita, Thai Duong, Siddhartha Jayanti, Cosmos Nicolaou,
 Himabindu Pucha, David Presotto, Adam Sadovsky, Suharsh Sivakumar, Gautham Thambidorai,
Robin Thellend for their contributions to designing and implementing the Vanadium authorization model. 
We are grateful to Mart{\'i}n Abadi and Mike Burrows for helpful comments on drafts of this tutorial.

\bibliographystyle{splncs03} \bibliography{references}

\begin{thebibliography}{10}
\providecommand{\url}[1]{\texttt{#1}}
\providecommand{\urlprefix}{URL }

\bibitem{fridge-attack}
{Fridge sends spam emails as attack hits smart gadgets}.
  \url{http://www.bbc.com/news/technology-25780908}

\bibitem{jeep-attack}
{Hackers remotely kill a jeep on the highway? with me in it}.
  \url{https://www.wired.com/2015/07/hackers-remotely-kill-jeep-highway/}

\bibitem{openid}
Openid. \url{http://openid.net/}

\bibitem{smart-meter-attack}
{Smart meters can be hacked to cut power bills}.
  \url{http://www.bbc.com/news/technology-29643276}

\bibitem{wired-iot-report}
{The Internet of Things is wildly insecure? and often unpatchable}.
  \url{https://www.schneier.com/essays/archives/2014/01/the_internet_of_thin.html}

\bibitem{Vanadium}
{Vanadium}. \url{http://vanadium.github.io/}

\bibitem{vanadium-auth-protocol}
{Vanadium authentication protocol}.
  \url{https://vanadium.github.io/designdocs/authentication.html}

\bibitem{ABPSST15}
Abadi, M., Burrows, M., Pucha, H., Sadovsky, A., Shankar, A., Taly, A.:
  Distributed authorization with distributed grammars. In: Programming
  Languages with Applications to Biology and Security. pp. 10--26 (2015)

\bibitem{AF99}
Appel, A., Felten, E.: Proof-carrying authentication. In: CCS. pp. 52--62
  (1999)

\bibitem{BPETVL14}
Birgisson, A., Politz, J.G., Erlingsson, U., Taly, A., Vrable, M., Lentczner,
  M.: Macaroons: Cookies with contextual caveats for decentralized
  authorization in the cloud. In: NDSS (2014)

\bibitem{BFI99}
Blaze, M., Feigenbaum, J., Ioannidis, J.: {The KeyNote trust-management system
  version 2}. RFC 2704 (Sep 1999), \url{https://tools.ietf.org/html/rfc2704}

\bibitem{BFL96}
Blaze, M., Feigenbaum, J., Lacy, J.: Decentralized trust management. In: IEEE
  Symposium on Security and Privacy. pp. 164--173 (1996)

\bibitem{BB02}
Borisov, N., Brewer, E.: Active certificates: A framework for delegation. In:
  NDSS. pp. 30--40 (2002)

\bibitem{BJ06}
Braz, C., Robert, J.: Security and usability: The case of the user
  authentication methods. In: Conference on L'Interaction Homme-Machine. pp.
  199--203 (2006)

\bibitem{CK02}
Canetti, R., Krawczyk, H.: Security analysis of {IKE}'s signature-based
  key-exchange protocol. In: {CRYPTO}. pp. 143--161 (2002)

\bibitem{CEEFMR99}
Clarke, D., Elien, J., Ellison, C., Fredette, M., Morcos, A., Rivest, R.:
  Certificate chain discovery in {SPKI/SDSI}. In: Journal of Computer Security.
  vol.~9, pp. 285--322 (2001)

\bibitem{DR08}
Dierks, T., Rescorla, E.: {The Transport Layer Security ({TLS}) Protocol
  Version 1.2}. RFC 5246 (Aug 2008), \url{https://tools.ietf.org/html/rfc5246}

\bibitem{EFLRTY99}
Ellison, C., Frantz, B., Lampson, B., Rivest, R., Thomas, B., Ylonen, T.: {SPKI
  Certificate Theory}. RFC 2693 (Sep 1999),
  \url{https://www.ietf.org/rfc/rfc2693.txt}

\bibitem{hp-iot-security}
Enterprise, H.P.: {Internet of things research study}.
  \url{http://www8.hp.com/h20195/V2/GetPDF.aspx/4AA5-4759ENW.pdf}

\bibitem{G89}
Gong, L.: A secure identity-based capability system. In: IEEE Symposium on
  Security and Privacy. pp. 56--63 (1989)

\bibitem{H12}
Hardt, E.: {The OAuth 2.0 Authorization Framework}. RFC 6749 (Oct 2012),
  \url{https://tools.ietf.org/html/rfc6749}

\bibitem{LABW91}
Lampson, B., Abadi, M., Burrows, M., Wobber, E.: Authentication in distributed
  systems: Theory and practice. In: SOSP. pp. 165--182 (1991)

\bibitem{NFG99}
Li, N., Feigenbaum, J., Grosof, B.N.: A logic-based knowledge representation
  for authorization with delegation. In: CSFW. pp. 162--174 (1999)

\bibitem{MAMGA08}
Myers, M., Ankney, R., Malpani, A., Galperin, S., Adams, C.: {X.509 Internet
  Public Key Infrastructure Online Certificate Status Protocol - OCSP}. RFC
  2560 (Jun 1999), \url{https://www.ietf.org/rfc/rfc5280.txt}

\bibitem{N93}
Neuman, B.C.: Proxy-based authorization and accounting for distributed systems.
  In: ICDCS. pp. 283--291 (1993)

\bibitem{baby-monitor-security}
Rapid7: {Hacking IoT: A case study on baby monitor exposures and
  vulnerabilities}.
  \url{https://www.rapid7.com/resources/iot/baby-monitors.jsp} (2015)

\bibitem{RL96}
Rivest, R.L., Lampson, B.: {{SDSI} - A simple distributed security
  infrastructure}. http://people.csail.mit.edu/rivest/sdsi11 (1996)

\bibitem{SFBHP08}
Santesson, S., Farrell, S., Boeyen, S., Housley, R., Polk, W.: {Internet X.509
  Public Key Infrastructure Certificate and Certificate Revocation List (CRL)
  Profile}. RFC 5280 (May 2008), \url{https://www.ietf.org/rfc/rfc5280.txt}

\bibitem{S13}
Schneider, F.B.: Untitled textbook on cybersecurity. chapter 9:
  Credentials-based authorization. \url{http://www.cs.cornell.edu/fbs/
  publications/chptr.CredsBased.pdf} (2013)

\bibitem{WT99}
Whitten, A., Tygar, J.D.: Why {Johnny} can't encrypt: A usability evaluation of
  pgp 5.0. In: USENIX Security Symposium. pp. 169--183 (1999)

\bibitem{WTSB16}
Wu, D.J., Taly, A., Shankar, A., Boneh, D.: Privacy, discovery, and
  authentication for the {Internet of Things}. Available at
  \url{https://arxiv.org/abs/1604.06959} (2016)

\bibitem{Z95}
Zimmermann, P.R.: The official PGP user's guide. MIT Press (1995)

\end{thebibliography}

\end{document}